 \documentclass[preprint2]{aastex}

\shorttitle{New lunar chronology}
\shortauthors{Marchi et al.}

\begin{document}


\title{A new chronology for the Moon and Mercury}


\author{Simone Marchi}
\affil{German Aerospace  Center (DLR), Institute  of Planetary Research,
     Rutherfordstr. 2, D-12489 Berlin}
\affil{ Dipartimento  di  Astronomia,   Universit\`a  di  Padova,  Vicolo
     dell'Osservatorio      2,       I-35122      Padova}
\email{simone.marchi@unipd.it}

 \author{Stefano Mottola}
\affil{German Aerospace  Center (DLR), Institute  of Planetary Research,
     Rutherfordstr. 2, D-12489 Berlin}

 \author{Gabriele Cremonese}
\affil{INAF, Osservatorio Astronomico di Padova, Vicolo
     dell'Osservatorio      3,       I-35122      Padova}

 \author{Matteo Massironi}
\affil{Dipartimento di Geoscienze, Universit\`a  di  Padova, via Giotto 1, I-35137, Padova}

\and

  \author{Elena Martellato}
 \affil{INAF, Osservatorio Astronomico di Padova, Vicolo
     dell'Osservatorio      3,       I-35122      Padova}

\begin{abstract}

In this  paper we present a new  method for dating the  surface of the
Moon,  obtained  by  modeling  the  incoming  flux  of  impactors  and
converting  it into  a  size distribution  of  resulting craters.   We
compare the results  from this model with the  standard chronology for
the Moon showing their similarities and discrepancies.  In particular,
we  find indications  of  a  non-constant impactor  flux  in the  last
500~Myr and also discuss the implications of our findings for the Late
Heavy Bombardment hypothesis.  We also show the potential of our model
for accurate dating of other inner Solar System bodies, by applying it
to Mercury.

\end{abstract}

\keywords{ solar system: general ---  planets and satellites: Earth,
  Mercury, Moon}

\section{Introduction}

Craters are among  the most spectacular surface features  of the solid
bodies of  the Solar System.  Cratering studies  provide a fundamental
tool for  the age determination of planetary  and asteroidal terrains.
Since the  beginning of the  lunar exploration, age estimates  for the
lunar  terrains were derived,  followed by  chronology models  for the
other terrestrial  planets.  The  development of the  lunar chronology
greatly helped in  interpreting the evolution of the  Solar System and
in particular  of our  own planet, the  Earth.  Recently, thanks  to a
fleet  of new  space  missions  (Mars Express  to  Mars, MESSENGER  to
Mercury, and  Kaguya to the Moon, only  to name a few),  this field of
research entered  a new exciting  phase, where accurate  age estimates
provide means for detailed small-scales geological studies.\\
The  method currently used  for dating  purposes defines  a chronology
(crater surface  density vs.  absolute  age) for a reference  body for
which radiometric ages are available for different terrains.  Then, on
the basis  of models  predicting the impactor  flux ratio  between the
reference body  and another generic  body, it is possible  to estimate
the age of the latter.  This method can be defined {\it experimental},
since  it develops  a chronology  of a  reference body  for  which two
measurable  quantities  are available:  absolute  ages  and number  of
craters for  selected areas.  The Moon  is the only body  of the Solar
System that could  fulfill both requirements so far.   This method has
been developed  and refined in the  last 30 years  by many researchers
and represents the  reference for dating purposes \cite[e.g.][]{har81,
neu83}.  In this  approach, the development of the  chronology for the
reference  body does  not  depend  explicitly on  the  physics of  the
cratering process.  However, such information becomes necessary -via a
cratering scaling law- in order  to apply the Moon chronology to other
bodies or to  infer the flux of impactors  from the observed cratering
record.\\
A possible  alternative -purely  {\it theoretical}- approach  could be
based on  the accurate estimation of the  time-dependent impactor flux
for each target  body, through the history of  the Solar System.  With
this method, a direct  comparison between the observed distribution of
craters on  a given terrain and  the isochrones produced  by the model
would  give its  age.  Although  the  knowledge of  the formation  and
evolution of the  Solar System greatly improved in  recent years, such
ambitious  goal  is  presently  far  beyond the  capabilities  of  the
available models, due  to the large uncertainties in  the early stages
of its formation.\\
In  this paper  we  introduce  a third  approach  for determining  the
chronology of  objects in the  inner Solar System. This  method, which
can be  considered as a {\it  hybrid} of the first  two approaches, is
based on  the dynamical model  by \cite{bot00, bot02,  bot05a, bot05b}
which describes the formation and  evolution of asteroids in the inner
Solar System.  In  this framework, first the flux  of impactors on the
Moon is derived, then the impactor size distribution is converted into
a cumulative  crater distribution via an appropriate  scaling law, and
finally the radiometric ages of different regions -the Apollo and Luna
landing sites- are  used for the calibration of  the lunar chronology.
The main advantage of our approach is that the adopted model naturally
describes the  distributions of the impactor sizes  and velocities for
any body in the inner Solar System. Thus, the lunar calibration can be
exported with precision to any  other body.  A drawback of the method,
however, is that the chronology of the reference body depends -through
the scaling law- on the physics  of the cratering process, which is as
yet  not fully  understood.   Furthermore, the  precision  of the  age
estimate   relies   on   the   accuracy  of   the   dynamical   model.
Encouragingly, the  adopted dynamical model is capable  of providing a
good  representation  of  the  present asteroid  population  and  size
distribution both in the near-Earth space and in the main belt, with a
maximum deviation amounting to less than a factor of 2 \citep{bot05b}.

\section{Modeling the cratering on the Earth-Moon system}

The present  inner Solar System is  continuously reached by  a flux of
small bodies having sizes smaller than  few tens of km.  A fraction of
these  bodies (namely  those having  perihelion $<1.3$~AU)  are called
near-Earth objects (NEOs).  Contributors  to this flux are represented
by main belt asteroids (MBAs)  and Jupiter family comets (JFCs).  Both
contributions  have been  modeled by  a number  of authors  and  it is
generally accepted  that the MBAs  constitute the main source  for the
flux    presently    observed    in    the    inner    Solar    System
\citep[e.g.][]{mor02}, therefore we neglect the cometary contribution.
This  flux  is  sustained by  a  few  fast  escape tracks,  which  are
continuously replenished  with new bodies  as a result  of collisional
processes and semimajor axis mobility.  The main gateways to the inner
Solar System are the $\nu_6$ secular resonance with Saturn and the 3:1
mean motion  resonance with Jupiter \citep{mor98, mor02}.   One of the
most  recent and accurate  models concerning  NEOs formation  has been
developed by \cite{bot00,  bot02}, while \cite{bot05a, bot05b} modeled
the main belt dynamical evolution.   These models have been adopted in
order to estimate the properties of the impactor flux at the Moon.

\subsection{Modeling the impactor flux: the stationary case}

\begin{figure}
\includegraphics[angle=-90,scale=.35]{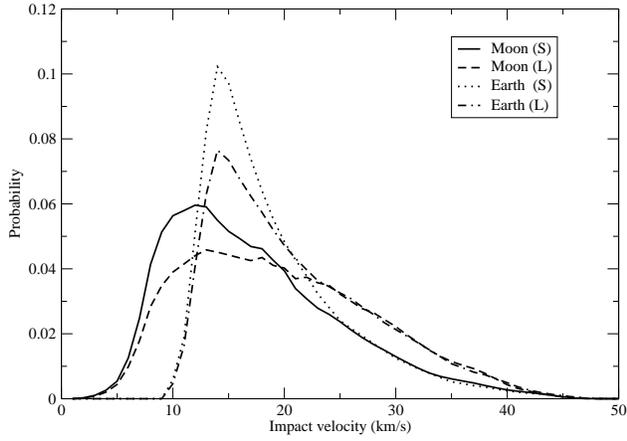}
\caption{Distributions  of impact velocities  on the  Moon and  on the
  Earth,  for two  different  impactor sizes.   The  smallest (S)  and
  largest (L) impactor  sizes used in these simulations  are 0.1~m and
  72~km,  respectively.  The  average impact  velocities are  18.6 and
  20.0~km/s,   respectively.   Notice   that   these  average   impact
  velocities are slightly lower then those used by \cite{neu94}, after
  applying the correction for  the average impacting angle of $\pi/4$.
  The distributions are  quite spread and therefore a  large number of
  impacts occur at velocities considerably different from the average,
  thus affecting the final crater distribution (see text). }
\label{moon_velocity}
\end{figure}

The flux of impactors may be described as a differential distribution,
$\phi(d,v)$, which  represents the number of incoming  bodies per unit
of  impactor  size ($d$)  and  impact  velocity  ($v$).  An  important
quantity -since it  can be constrained from observations-  is the size
distribution of  the incoming bodies,  namely the number  of impactors
per  unit impactor  size.  Let  $h(d)$ be  such distribution,  we then
have:  $h(d)\equiv\int  \phi(d,v)\,dv$.   Therefore, without  loss  of
generality, we can write:

\begin{equation}
\phi(d,v)=h(d)f(d,v) \label{phi}
\end{equation}

where $f(d,v)$ is the distribution  of the impact velocities (i.e. the
impact probability  per unit impact  velocity) for any  dimension $d$,
and  which  fulfills  the  normalization constraint  $\int  f(d,v)\,dv
\equiv 1$. \\
We used  the NEO orbital  distribution model computed  by \cite{bot00,
bot02},  generated by integrating  two sets  of test  particles placed
into  the   $\nu_6$  and   3:1  resonances  (with   $2\times10^6$  and
$7\times10^5$  particles,  respectively).    For  each  particle  they
computed the  impact probability and  the impact velocity  relative to
the target body.  Following the  work of \cite{mar05}, we computed the
impact   velocity  distribution   (corrected  for   the  gravitational
cross-section  of the  targets).  We  assumed that  the  Moon revolves
around  the  Sun  with  an  orbit  identical to  that  of  the  Earth.
Moreover, we  corrected the impact probability provided  by the Bottke
et  al  model  in  order  to account  for  the  different  collisional
life-time  of the particles.   This is  necessary because  the orbital
evolution  of the  particles is  coupled with  their size  through the
relationship  $\tau  \propto  \sqrt{d}$  \citep{far98,  bot05b},  with
$\tau$  being   the  particles'  collisional   life-time.   In  figure
\ref{moon_velocity}   we   show   the  resulting   impactor   velocity
distribution  $f(d,v)$ for  the  two  extreme size  bins  used in  our
simulations and  for both the Moon and  the Earth\footnote{The overall
shape   of   the   distributions   vaguely  resembles   a   Maxwellian
distribution.  It  is interesting that the distributions  have a large
number   of  high   velocity  (i.e.    $>20$~km/s)   impactors.   This
distribution  may  help  to  explain  some anomalies  found  for  some
terrestrial impact  craters.  In  particular, only 3  out of  about 60
known craters with  ages $<60$~Myr have been at  present associated to
tektite  strewn  fields  \citep{fre98}.   Since  tektites  are  highly
shocked  super-heated  melts   ejected  during  hypervelocity  impacts
\cite[e.g.][]{gla90, koe96}, from  simple statistical arguments we may
estimate that  a threshold perpendicular  impact velocity of  at least
24~km/s is required for their formation (more than 33~km/s modulus for
a $\pi/4$ impacting angle).}.\\
The impactor size distribution for the Earth ($h_e$) can be written as
follows:

\begin{equation}
h_e(d)=P_eh_{n}(d) \label{he}
\end{equation}

where $h_{n}$ is  the NEO differential size distribution  and $P_e$ is
the  intrinsic  collision  probability.  Following  \cite{bot05b},  we
adopted  $P_e=2.8\cdot10^{-9}$~yr$^{-1}$.  The distribution  $h_n$ has
been derived from  \cite{bot02}, and it can be scaled  to the Moon and
other bodies.  In the case of  the Moon, this is  done considering the
Earth-Moon ratio of  the impact probability as a  function of impactor
size  \citep{mar05}.  Let  $\Sigma(d)$ be  this ratio,  then  the Moon
impactor size distribution becomes:

\begin{equation}
h_m(d)=P_eh_{n}(d)\Sigma(d) \label{sigma}
\end{equation}

\begin{figure}
\includegraphics[angle=-90,scale=.50]{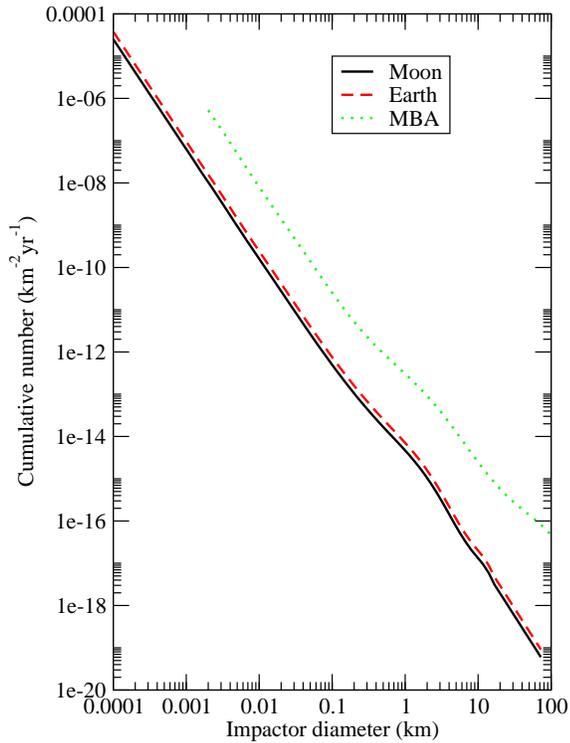}
\caption{Model cumulative size distributions  of impactors on the Moon
and on the Earth  \citep[after][]{bot02}.  An arbitrarily rescaled MBA
cumulative  distribution  \citep[after][]{bot05a}  is also  shown  for
comparison.  The  main belt distribution shows an  S-shaped feature in
the  0.2-2~km   size  range.   Also  the  NEO   distribution  shows  a
qualitatively similar  feature.  The two  distributions, however, have
quantitatively  different shapes, due  to selection  mechanisms taking
place  during  the dynamical  transport  from  the  main belt  to  the
near-Earth space.}
\label{moon_size}
\end{figure}

In  figure \ref{moon_size} the  cumulative impactor  size distribution
for both  the Moon and  the Earth are  shown.  A similar  approach can
also be  used for evaluating the  flux on other  targets, like Mercury
(see section  \ref{mercu}).  In  equation \ref{phi} the  dependence of
$h$  and  $f$  over time  is  neglected  since  we are  considering  a
stationary  problem,  while  eq.   \ref{he} and  \ref{sigma}  strictly
describe the present flux.  Arguments about possible variations in the
flux intensity and  shape have been discussed since  the late 70s, and
this still  remains one  of the  most debated topics  in the  field of
lunar    cratering.    In   their    work,   Neukum    and   coworkers
\cite[e.g.][]{neu83, neu94}  generally assume a constant  shape of the
impactor  size distribution  over time,  while others  claim  a sudden
change  in the  shape after  the Late  Heavy Bombardment  (LHB), about
3.8~Gyr ago \citep[e.g.][]{str05}.  One of the potentials of our model
is that we can implement  and evaluate the effects of a time-dependent
size-distribution on  the crater cumulative distribution,  as shown in
section \ref{t-mpf}.

\subsection{The scaling law}\label{scalinglaw}

\begin{figure}
\includegraphics[angle=-90,scale=.35]{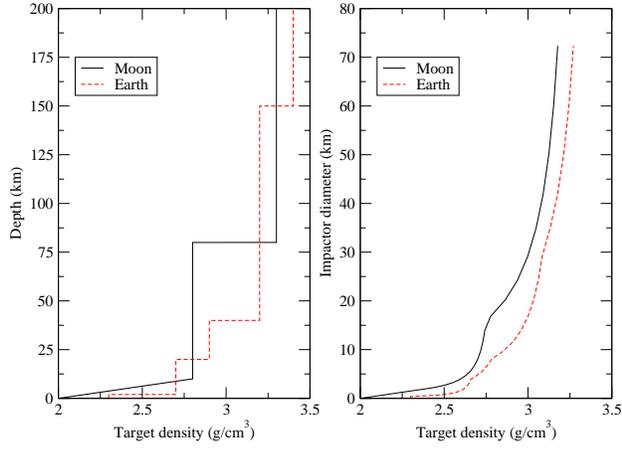}
\caption{Assumed  density profiles for  the Moon  and the  Earth.  Left
  panel: lunar and terrestrial density profiles vs depth. Right panel:
  average  density vs  impactor  size. The  average  density has  been
  obtained by  averaging the density up  to a depth  of 10$\times$ the
  impactor  radius.  A  similar consideration  holds for  the strength
  (see text for details).}
\label{density}
\end{figure}

In  order to convert  the flux  of impactors  into a  resulting crater
distribution, we need to apply a so-called crater-scaling law. Such
scaling  law attempts  to describe  the outcome  of a  cratering event
based on the impact energy and  on the physical properties of both the
target and the projectile.  Once an appropriate scaling law is chosen,
we  may write  the final  crater  diameter $D$  as a  function of  the
impactor  diameter,  impact  velocity,  and  a  number  of  parameters
(density,  strength...   indicated   by  $\vec{p}$),  namely  $D={\cal
S}(d,v,\vec{p})$.   Despite the great  effort (both  computational and
experimental) devoted to this task, the physics of crater formation is
far from being completely understood.   For this reason, we decided to
explore how the predicted crater distribution depends on the choice of
the scaling  law.  For  this purpose we  considered the  three scaling
laws most  used in  this field.  The  first one by  \cite{hol07} (H\&H
hereinafter), reads:
 
\begin{equation}
D=kd\biggl[\frac{gd}{2v_{\perp}^2}\biggl(\frac{\rho}{\delta}\biggr)^\frac{2\nu}{\mu}+\biggl(\frac{Y}{\rho v_{\perp}^2}\biggr)^\frac{2+\mu}{2}\biggl(\frac{\rho}{\delta}\biggr)^\frac{\nu(2+\mu)}{\mu}\biggr]^{-\frac{\mu}{2+\mu}} \label{hh}
\end{equation}

where $g$  is the target gravitational acceleration,  $v_\perp$ is the
perpendicular  component of  the  impactor velocity,  $\delta$ is  the
projectile  density,  $\rho$  and  $Y$  are the  density  and  tensile
strength of  the target.  The quantities  $k$ and $\mu$  depend on the
target  material and  $\nu$  on its  porosity.   H\&H estimated  these
latter  parameters by best-fitting  over a  range of  experiments done
with  different  materials. From  this  study  we  adopted the  values
$k=1.03$ and  $\mu=0.41$ for cohesive soils,  and $k=0.93$, $\mu=0.55$
for rocks.  We used $\nu=0.4$  in all cases (H\&H).  Equation \ref{hh}
accounts  for  the transition  between  strength  and gravity  regime,
allowing a  smooth transition  between those extreme  conditions.\\ 
The  second  scaling  law  considered  is  reported  by  \cite{iva01},
\cite[adapted from][]{sch87} (I\&S hereinafter):

 \begin{equation}
\frac{D}{d(\delta/\rho)^{0.26}v_\perp^{0.55}}=\frac{1.28}{[(D_{sg}+D)g]^{0.28}}
\end{equation}

where    $D_{sg}$   represents    the    strength-gravity   transition
crater. $D_{sg}$ has been set equal to 120~m and 30~m for the Moon and
the  Earth, respectively \citep{asp96}.\\  
Finally,  we have  considered  the scaling  law  used by  \cite{stu04}
\cite[adapted from][]{sho90} (S\&S hereinafter):

\begin{equation}
D=0.0224\biggl(W\frac{\delta}{\rho}\biggr)^{0.294}\biggl(\frac{g_e}{g}\biggr)^{1/6}(\sin{\alpha})^{2/3}
\end{equation}

where, $W$ is the impactor  kinetic energy, $g_e$ is the Earth gravity
and  $\alpha$  is the  impactor  angle  with  respect to  the  surface
(vertical  impacts  correspond   to  $\alpha=\pi/2$).   The  numerical
multiplicative factor takes into account the correction from transient
to final crater dimension \citep{stu04}.\\
The crater size  $D$ reported in these formulas  should be regarded as
the  size of  the final  crater.  However,  following  \cite{pik80}, a
correction for  the transition between simple and  complex craters has
also been applied to all the scaling laws in the form of:

\begin{eqnarray}
D & = & D \hspace*{2cm} {\rm if}  \;\; D < D_{\star} \\
D & = & \frac{D^{1.18}}{D_{\star}^{0.18}}  \hspace*{1.4cm} {\rm if}  \;\; D > D_{\star}
\end{eqnarray}

where $D_{\star}$  is the  diameter of the  transition from  simple to
complex crater\footnote{$D_{\star}$  has been  set equal to  18~km and
4~km  for the  Moon and  Earth, respectively  \citep{iva01}.   }.  All
impacts are assumed to occur at the most probable impact angle, namely
$\pi/4$ with respect to the normal to the surface.\\
All the scaling  laws assume that the density and  the strength of the
target are constant throughout  the body.  However, most target bodies
are characterized  by a layered  structure, with a  density increasing
with depth.  Therefore, in order to produce a better representation of
the target density, we have  computed the average density over a depth
of about 10  times the radius of the projectile.   This is because the
size of  the crater is  -on average over different  impact conditions-
about  a  factor  ten  larger  than  the size  of  the  impactor.   In
particular, in the  case of the Moon we have assumed  a 10-km layer of
fractured silicates (megaregolith and heavily fractured anorthosites),
on top of a bulk anorthositic crust in turn laying above a peridotitic
mantle   (see  fig.    \ref{density}).   For   the  Earth   (see  fig.
\ref{density})  we  considered  the following  lithospheric  structure
\citep{and07}:  a 2~km  thick  layer of  sedimentary  rocks, a  mainly
granitoid upper-crust down  to a depth of 20~km,  a denser lower crust
down  to 40~km,  a peridotitic  lithospheric mantle  which is  in turn
characterized  by   an  upper  layer   with  an  average   density  of
3.2~g/cm$^3$ (up to 150~km), and  a lower layer with an higher density
(3.3~g/cm$^3$).  The  impactor density  has been set  to 2.7~g/cm$^3$.
Regarding the  strength, the value  reported by \cite{asp96}  for bulk
silicates, namely $Y_0=2\times10^8$~dyne/cm$^2$  was used both for the
Moon and  Earth crusts.  However, we  assume a linear  increase of the
strength, from  zero at the surface up  to $Y_0$ at the  bottom of the
heavily fractured layer for the Moon in order to take into account the
cohesionless  regolith  layer  on   the  surface  and  the  underlying
megaregolith  and anorthosites  which  are likely  characterized by  a
progressive  decrease  in  fracture  density with  depth.   A  similar
assumption was adopted also for  the layer of sedimentary rocks on the
Earth.    The   strength  at   the   Earth's   surface   was  set   to
0.3~$\times10^8$~dyne/cm$^2$.\\
Following  our description  of the  Moon crust,  its upper  layers are
mainly made of highly  fractured materials, which behave like cohesive
soils.  Therefore, we set a  sharp transition in the scaling law, from
cohesive soil in  the case of small impactors to  hard rock for larger
ones,  at a projectile  size of  1/20$^{th}$ of  the thickness  of the
heavily fractured  silicate layer (i.e.   0.5~km). For the  Earth only
the hard rock  scaling law was used since only  rocky layers have been
assumed at the surface.

\section{ The Model Production Function (MPF)}

We  have now  all the  necessary inputs  for computing  the cumulative
crater  size distribution  (hereafter the  Model  Production Function,
MPF) for the  target bodies. Let $\Phi(D)$ be  the crater differential
distribution, we may write:

\begin{equation}
{\rm MPF}(D)=\int_{D}^\infty \Phi(\tilde{D})d\tilde{D}
\end{equation}

where $\Phi$ can  be expressed in terms of $\phi$  and ${\cal S}$ (see
appendix for  details).  In figure \ref{moon_scaling}  our lunar model
production function is shown for different scaling laws along with the
Neukum production function  for the Moon (NPF).  The  NPF has been the
traditional reference for  dating purposes since the late  70s, with a
few  revisions   in  more  recent  times   \citep{neu83,  neu01}.   An
alternative  production function,  the  so-called HPF,  has also  been
proposed  by \cite{har81}.   A detailed  comparison between  these two
production functions can be  found in \cite{neu01}.  For our purposes,
we recall that  the NPF shows an overall agreement  with the HPF, even
if  some  discrepancies  are  present.   Nevertheless,  as  stated  in
\cite{neu01},  the NPF-based  chronology represents  the  current best
interpretation  for the lunar  cratering chronology.   Moreover, since
the NPF extends to a wider size range than the HPF, the former is more
suitable for a  thorough comparison with the MPF.   For these reasons,
here we provide  detailed comparison between the MPF  and the NPF.  In
the  following paragraph,  we  briefly  recall how  the  NPF has  been
derived.\\
There are several versions of the  NPF.  In this paper we refer to the
{\it old}  NPF \citep{neu94} and  to the {\it new}  NPF \citep{neu01}.
In  both cases, the  underlying assumption  is that  the shape  of the
production function  has been constant, with only  the absolute number
of  craters changing  over  time.   Therefore, the  NPF  was built  by
vertically re-scaling  crater counts  from areas with  different ages,
until  they  overlapped. The  final  curve  was  then expressed  as  a
polynomial fit  of the  re-scaled data points.   The strength  of such
procedure is that it is  not model dependent because based on measured
data.  On  the other hand this  procedure, relying on  the accuracy of
the  measurements  in multiple  overlapping  regions, is  particularly
prone to  severe error propagation,  especially at the  large diameter
end.   In addition, in  some regions  the cumulative  distributions of
craters  can  be altered  by  surface  processes  like sporadic  magma
effusions, ejecta  from other craters etc.  The  subsequent effects of
superposition  of geological  units  of different  ages, normally  not
recognizable on remote  sensing images, can lead to  mixed crater size
distributions \citep{neu76}.
Figure  \ref{moon_scaling}  clearly shows  how  the  change of  crater
  counting on single regions may alter the final shape of the NPF.  In
  particular, the  discrepancies between the  old and the new  NPF are
  due  to an  adjustment  of  the crater  counts  for Orientale  Basin
  \citep{iva01}.
A few considerations can  be drawn from fig.  \ref{moon_scaling}.  The
first one  is that different  scaling laws produce differences  on the
MPF as large as  a factor of ten for large craters.   We find that the
MPF produced with  the S\&S scaling law is not  in good agreement with
the  NPF throughout  the whole  size  range.  When  applying the  same
scaling  law to  the  observed NEO  population,  \cite{stu04} found  a
better match to the new NPF for craters $>10$~km.  This discrepancy is
mainly due  to the different  value of the intrinsic  probability that
Stuart     and    Binzel    used     in    their     model    (namely,
$1.50\cdot10^{-9}$~yr$^{-1}$).   However,  we  believe the  Bottke  et
al. determination  of the intrinsic  collision probability to  be more
accurate since  it relies on  a better treatment of  the gravitational
focusing  effects (A.   Morbidelli, pers.   communication).   The main
difference between the  H\&H and the I\&S scaling  laws takes place at
crater  sizes smaller  than 10~km.   This can  be traced  back  to our
choice of using  the cohesive soil scaling law  for small crater sizes
in the H\&H  scaling law which causes an inflection  point in the MPF.
This  transition  in  the  target's  crustal  properties  produces  an
S-shaped feature in the MPF, which is qualitatively similar to the one
found  in the  NPF.   This  finding would  suggest  that the  observed
S-shaped feature  in the NPF is  caused by physical  properties of the
target, rather than  reflecting the shape of the  size distribution of
the   impactors,  as  proposed   by  \cite{wer02}   and  \cite{iva02}.
Actually, a weak wavy feature in the relevant size range is present in
the $h_{n}(d)$ which, however, is smeared out when passing to the MPF.
However, in order  to finally assess this issue,  a better estimate of
the observed impactor size distribution for sizes $<0.1$~km is needed.
It  should  also be  mentioned  that the  detailed  shape  of the  NPF
somewhat depends  on the fitting  procedure used to construct  the NPF
itself, as shown by the mismatch between the old and new NPF for large
craters.\\
%
%
%
In  conclusion,  despite  of  all  the  caveats  mentioned,  it  is  a
remarkable result that the present  MPF obtained with the H\&H scaling
law and the NPF (both new and  old) are similar within a factor of two
throughout the whole size range  considered, that is from 0.01~km to a
few hundred  km.  On the  other hand, the  S\&S and I\&S  scaling laws
fail to accurately reproduce the absolute number of craters per km$^2$
per  yr,  with a  maximum  deviation  of a  factor  of  10.  The  good
agreement of the  absolute density of craters per  yr derived from the
observations (either  the NPF or  the HPF) and  the MPF with  the H\&H
scaling law  is therefore a  clear indication that the  latter scaling
law is the  more appropriate in describing the  lunar cratering.  This
agreement is  particularly interesting,  considering that the  NPF and
MPF are  derived from completely independent methods:  one being based
on  crater counting,  and the  other being  the result  of theoretical
modeling.   For these reasons,  we decided  to restrict  the following
analysis to the H\&H scaling law.  We refer to this MPF as to the {\it
nominal model}.\\

\subsection{The reliability of the nominal MPF}

In this  section we  explore the effects  that the  various parameters
involved in  our model have on  the nominal MPF described  in the last
section.   In particular,  we analyze  the scaling  law, the  NEO size
distribution and the assumed density and strength profiles.\\
The H\&H scaling law depends on three parameters (namely $k, \nu$ and
  $\mu$) which have  been derived by best fit  to laboratory data.  We
  applied  a  variation  of  10\%  around their  nominal  values,  and
  computed the  resulting MPF.  It results that  $k$ and $\nu$  have a
  negligible effect on the MPF.   Changing $\mu$ of $\pm$10\%, the MPF
  shifts vertically of about a factor of $\pm2$, respectively.\\
 A  detailed   comparison  of  the  modeled  and   observed  NEO  size
distributions can be found  in \cite{bot05a,bot05b}. For our purposes,
we  notice that they  basically overlap  for $d>1.5$~km.   For smaller
impactor sizes,  the situation is  less clear since  different surveys
produce  slightly  different observed  distributions.  Anyway, we  can
safely state that we have a maximum deviation between the observations
and  the model  distributions  of about  a  factor of  1.8 (being  the
observations lower than the model).\\
The scaling law is weakly dependent on the density, so a change in the
numerical values  affects the  MPF in a  negligible way.   Notice also
that  craters larger  than about  0.1-0.2~km form  in  gravity regime,
hence the rock strength is  not important at those sizes.  For smaller
sizes, the  shape of the  MPF may depend  on the local  strength.  The
density and  strength profiles  adopted in this  work are meant  to be
average profiles for the Moon. Local variations to the mean values may
slightly change the MPF at small crater sizes.  \\
A more  important parameter  is the assumed  depth for  the transition
from cohesive  to hard  rock scaling law.   A variation in  this value
causes a  shift of the  inflection point in  the MPF within  the range
$1<D<10$~km.

\section{A new chronology for the Moon}

The MPF  can be used as  a reference to develop  the lunar chronology.
For this purpose,  we used literature crater counts  from regions with
known radiometric ages as calibration  regions.  A list of the regions
used is reported in table \ref{table1}.\\
In order to obtain the lunar chronology, we first assumed that the MPF
shape  remained constant over  time.  For  each calibration  region we
determined the proportionality  factor which gave the best  fit to the
data by  minimizing the $\chi^2$.   Finally, we determined  the crater
cumulative number at 1~km ($N_1$).
The fitting  procedure used is particularly important,  as it directly
affects the chronology. For this reason, we performed several tests in
order to assess  its stability.  A major aspect  consisted in choosing
the weights given to the  measured data points.  It is common practice
in  crater   counting  to  assign   each  data  point  an   error  bar
corresponding to  the square root of  the number of  counts, under the
assumption  that  the  measurements  follow  a  Poisson  distribution.
However,  in the  case  in which  significant  systematic effects  are
present, this procedure leads  to underestimate the total error.  This
problem is  particularly evident when dealing with  small craters that
are  at the limit  of the  image resolution.   Because of  their large
number,  the  statistical error  for  these  craters is  comparatively
small.   On  the other  hand,  due  to  the difficulty  of  positively
identifying craters at the limit of the resolution, these measurements
could be affected  by a significant bias.  This  effect may affect the
$N_1$ values  up to a factor  of 3-4 in some  regions and consequently
have a non negligible effect  on the chronology.  After several tests,
we decided to use the following weights:

\begin{equation}
\sigma_t^2=N+\kappa N^2
\end{equation}

where  $\kappa$  was  set  to  0.5.   In this  way,  the  total  error
($\sigma_t$)  for  points  with  very low  statistical  $\sigma$  were
increased, while those with  large statistical error (i.e.  low number
of craters) were basically left unchanged.\\
In order  to validate this  procedure, we applied  it to the  old NPF,
being able  to reproduce the  \cite{neu94} results within  their error
bars\footnote{Here we used the old  NPF instead of the new NPF because
even the latest chronology by \cite{neu01} uses the old NPF.}.

\subsection{Moon cratering}

In this  section we  present the  results of the  fits applied  to the
regions   used   to  calibrate   the   lunar   chronology  (see   tab.
\ref{table1}).  The  fitting procedure was successful for  most of the
regions.  In  several cases a very  good fit was  achieved through the
measured range of the  cumulative distributions, indicating an overall
agreement between  the shape of the  MPF and the  data.  However, some
discrepancies occurred for the highlands (for $D>100$~km), Mare Crisum
and  Mare Tranquillitatis.   In the  case of  the highlands  (see fig.
\ref{highlands}), the  discrepancy seems to suggest  a different shape
of the impactor size distribution  in the past, with a larger fraction
of  medium-to-large  impactors (see  section  \ref{t-mpf} for  further
details).   In  the  case  of  Mare Crisum  and  Mare  Tranquillitatis
regional phenomena can be responsible for the observed deviations.  In
particular,  in  both  cases  several  consecutive  magma  flows  have
partially  covered  the population  of  small  craters, affecting  the
observed  cumulative distribution  \citep{neu76,  boy77}.  Indeed  the
absolute   radiometric   ages   derived   from   A11   samples   (Mare
Tranquillitatis), span from  3.5 to 3.85~Gyr, whereas the  ages of the
Luna  24 basalts  (Mare Crisium)  are  clustered into  at least  three
different  peaks  at  2.5,  3.3, 3.6~Gyr,  respectively  \citep{bir78,
sto01, fer05}.  In presence of such surface phenomena, the fit must be
constrained  on  the  basis   of  geological  considerations,  as  the
automatic fit may be misleading.  In particular, as formerly suggested
by \cite{neu76},  the crater size-frequency  distribution derived from
these  regions leads  to a  composite curve  in which  smaller craters
reflect the younger radiometric ages, while the larger ones correspond
to older  ages.  Thus, for Mare  Tranquillitatis we forced  the MPF to
overlap  the  cumulative curve  at  0.5~km  for  the age  of  3.55~Gyr
(referred to  as {\it young}), and  at 0.9~km for the  age of 3.72~Gyr
(referred to as {\it old}).  Similar considerations hold also for Mare
Crisum where the MPF was fitted to craters with diameters in the range
1 to 2~km (details on the  fit of the calibration regions are reported
in the online material).\\
Another effect  influencing the accuracy  of the crater  statistics is
the  possible  presence  of  secondary craters.  Even  though  authors
publishing   crater   counts   pay   particular   attention   to   the
identification  and  rejection of  secondary  craters,  in some  cases
confusion with  primary craters is  difficult to avoid.   According to
\cite{iva06},  secondaries  are  likely  negligible on  young  regions
because they did not have the time to accumulate in large numbers.  On
the other  hand, older  regions, namely the  highlands and  maria, may
show   secondary   craters   caused   by  subsequent   large   impacts
\citep{wil76}.    Nevertheless,   \cite{neu94}   concluded  that   the
secondaries  are likely not  relevant.  For  the present  analysis, we
assumed  that   the  observed  crater   cumulative  distributions  are
resulting from the primary flux of impactors.

\subsection{Earth cratering}

Cratering studies  are much  more difficult on  the Earth than  on the
Moon, due to strong resurfacing processes and consequent alteration of
the  crater morphologies.  Nevertheless,  Earth craters  are important
because they can be studied  in detail and precisely dated.  Therefore
they  can be  used  to further  constrain  the flux  of impactors,  in
particular  for young  ages, that  are scarcely  represented  in lunar
data.  In  this work  two data sets  of terrestrial craters  have been
considered. The  first one consists  of all large  craters ($D>20$~km)
found on the  North American and Euroasiatic cratons,  as presented by
\cite{gri79}  (updated with  few new  findings from  the  Earth Impact
database\footnote{http://www.unb.ca/passc/ImpactDatabase/.}).
Following \cite{gri79}, we  have adopted for these two  cratons an age
of 0.375~Gyr.  Some  sectors of both cratons were  probably exposed to
the  meteoroid flux  already  from the  early Ordovician  (0.450~Gyr),
whereas others  only from 0.300~Gyr ago \citep{gri79}.   The choice of
0.375~Gyr is further  supported by the age of  the older craters that,
with only few exceptions, are around this value.  \\
The second  data set represents  a subset of  the first  one, including
only craters  with radiometric ages less then  0.120~Gyr.  This choice
has been made following \cite{gri93},  in order to overcome biases due
to  erosion  and burial  of  craters  on  the Earth's  surface.   Both
cumulative  distributions were  fitted using  our Earth  MPF,  and the
obtained  $N_1$ was  rescaled to  the  lunar case  by considering  the
gravitational  focusing   and  the  different   relationships  between
impactor   and   crater   size   on   the   two   bodies   (see   fig.
\ref{moon_earth}).  The  resulting values for $N_1$ are  shown in tab.
\ref{table1}, with more details supplied in the online material.

\subsection{MPF chronology}\label{chr}

The  lunar  chronology  has  been   developed  on  the  basis  of  the
relationship between  the derived $N_1$ values and  the absolute ages.
However,  this task  is not  trivial  since only  18 measurements  are
available, 12  of which  are older than  3~Gyr.  Moreover,  the region
from 1~Gyr to 3~Gyr lacks completely in data points.  Therefore a very
accurate chronology cannot be expected, in particular for ages younger
than  3~Gyr.    In  this  work   the  radiometric  ages   proposed  by
\cite{sto01} were used.\\
In order to describe  the lunar chronology, \cite{neu83} suggested the
following approximation:

\begin{equation}
N_1=a(e^{bt}-1)+ct. \label{chrono_eq}
\end{equation}

This function assumes a linear  relationship between $N_1$ and the age
($t$) for $bt\ll1$, while  for $bt\gg1$ $N_1$ increases exponentially.
Although  this  is  a  rather  simple  description  of  the  data,  it
accurately fits the  data points in the Neukum  chronology.  We fitted
the same function to our MPF-based data points, with the results shown
in   fig.   \ref{mpf_neukum_fit}.    The   derived  coefficient   are:
$a=1.23\times10^{-15}$,    $b=7.85$,    $c=1.30\times10^{-3}$.     For
comparison,  \cite{neu94} obtained:  $a=5.44\times10^{-14}$, $b=6.93$,
$c=8.38\times10^{-4}$,    which    are    quite   similar    to    our
result\footnote{As previously  discussed, old regions  with cumulative
counting  extending to sub-km  sizes may  be affected  by secondaries.
These   regions   are:  Descartes   Formation,   Mare  Imbrium,   Mare
Fecunditatis, Taurus Littrow Mare, Mare Tranquillitatis.  Therefore we
also  performed a  best  fit excluding  these  regions. The  resulting
coefficients        are        $a=3.07\times10^{-15}$,       $b=7.63$,
$c=1.33\times10^{-3}$, and the resulting chronology curve is basically
overimposed to the one obtained  using all the regions.  Therefore the
secondary craters -if presents in the cumulative distributions- do not
affect the  chronology.  }. The \cite{neu94} chronology  curve is also
plotted in fig. \ref{mpf_neukum_fit}. \\
While the  derived best fit is  in overall agreement  with the nominal
MPF-based data  points, there are some deviations  that mainly concern
young data points, which lie  systematically above the best fit curve.
It is  remarkable that in  our chronology, the proposed  equation fits
well all the  points older than Copernicus \cite[which  was an outlier
in the  chronology of][]{neu94}, while it underestimates  -by a factor
of    2-   all    the    points   younger    than   Copernicus    (see
fig. \ref{mpf_neukum_fit}).

\subsection{Non-constant flux in recent times?}

As mentioned, the MPF-based chronology is reasonably in agreement with
that  derived from  the  NPF.  The  discrepancies  observed for  young
regions  could be  due to  uncertainties  involved in  our model.   In
particular,  a concern regards  the slope  of the  impactor cumulative
size distribution for small  dimensions which is poorly constrained by
observations. Errors  in the  slope in this  size range may  result in
inaccurate estimates of $N_1$ for  the young lunar regions.  The slope
of the  NEO model  for impactor diameters  smaller than  0.1~km (which
roughly  corresponds to  1~km crater  size) is  -2.6, which  becomes a
slope of -3.1  in the MPF. This  slope is very similar to  that of the
NPF for  the same  size range,  which is equal  to -3.2.   This slight
difference in  slope can account  for half of the  observed difference
between  our and Neukum's  chronology.  We  underline that  such slope
variations may affect  the Cone, Tycho and North  ray craters, but not
the terrestrial crater counts  since they have $D>20$~km. Therefore it
is remarkable that terrestrial craters lie on the same linear trend of
the young lunar data  points (fig.  \ref{mpf_neukum_fit}, left panel).
The NEO population model adopted here is in overall agreement with the
available observed  data \cite[see fig.  16  of][]{bot05b}.  For small
dimensions, however, the real  NEOs size distribution is poorly known.
The   only   observational  constraints   are   the  small   fireballs
\citep{hal96}  and  bolides  \citep{bro02}  detected  in  the  Earth's
atmosphere.  A careful  check at fig.  16 of  \cite{bot05b} shows that
the NEO population model slightly  overestimates -by a factor of about
1.5- the predicted  number of NEOs from bolide  data.  This occurs for
the impactor size range responsible for the cratering observed for the
Cone,  North Ray  and  Tycho  craters.  Based  on  the data  presently
available, it is very  difficult to understand whether this difference
is real or rather due to the involved uncertainties.  Nevertheless, if
we maintain the present number of NEOs at $d=1$~km, we obtain that the
$N_1$  values corresponding  to  the younger  lunar  regions would  be
further increased.\\
Other  possible sources of  uncertainties may  arise from  the scaling
law.  Nevertheless,  variations in  the parameters produce  a vertical
shift of the  MPF and not a change of the  shape, therefore this shift
alone cannot affect the determination of $N_1$ values. The only way to
obtain a change in the $N_1$ values with respect to the nominal model,
is to introduce a slope variation  in the MPF for $D<1$~km. This would
be   the   case,  if   the   inflection   point   would  slide   below
$D\sim1$~km. This would  imply a very low thickness  for the fractured
layers which  seems improbable, according to our  understanding of the
lunar crust.  Nevertheless,  if this would be the  case, $N_1$ for the
young lunar regions would be further increased. \\
It is also possible that  the MPF-based $N_1$ values are accurate, and
the  simple  relationship  proposed  by  \cite{neu83,  neu94}  is  not
adequate to describe the data. This would be the case, for example, if
the impactor flux  had not been constant during  the past $\sim$3~Gyr.
Due to the lack  of data points in the range 1-3~Gyr,  we are not able
to study  in detail  possible variations in  the flux.   However, fig.
\ref{mpf_neukum_fit}  seems  to  suggest  that  a  recent  phase  (age
$<0.4$~Gyr) characterized by a  constant impactor flux was preceded by
a  phase (between  0.8 and  3~Gyr) with  a lower  and  nearly constant
impact rate. A single event  placed around 0.4~Gyr ago and lasting for
0.2-0.3~Gyr  that would  increase the  impactor flux  by a  factor 2-3
would explain the observed data.
%
Such a  recent change  in impactor flux  would also have  affected the
older regions, but here the relative increase in the number of craters
would be very  low compared to the accumulation  of craters over 3~Gyr
or  more, hence  this effect  would be  negligible.  This  scenario is
qualitatively in  agreement with the recent suggestion  of an increase
in  the number  of impactors  in  the inner  Solar System  due to  the
formation  of  dynamical  families   -as  a  results  of  catastrophic
disruption of  asteroids- in the  main belt \citep{nes07,  bot07}.  In
particular, the  Baptistina and Flora  families are estimated  to have
formed about  140~Myr and 500~Myr ago,  respectively.  Their proximity
to  the  resonances led  to  an  increase in  the  member  of the  NEO
population,  starting few  tens of  Myr  after their  formation, as  a
consequence  of  the Yarkowsky-effect-driven  decay  of the  semimajor
axis.  In  the case of Baptistina  the maximum flux  was reached about
60~Myr after  its formation (i.e.   80~Myr ago).  Therefore  this flux
may   have  affected   the   Cone,  North   Ray   and  Tycho   craters
counting\footnote{See  the   discussion  in  \cite{bot07}   about  the
probability  of  Tycho  itself  being  formed  by  an  impact  with  a
Baptistina fragment.}, and also the young terrestrial craters.\\
Concerning the  formation of  Flora, similar considerations  may hold.
In addition, the  Flora family has been connected to  the spike in the
infall  of  L-chondrite meteorites  that  occurred  about 470~Myr  ago
\citep{bog95}.  A  similar spike has  also been detected in  the lunar
glass  spherules and  -to  a  lesser extent-  in  lunar impact  clasts
\citep{cul00,  coh05}.   These spikes  may  have  affected mostly  the
terrestrial  cratering  and the  Copernicus  counting.  Although  such
enhancement  of  the flux  was  probably  formed  by sub-km  impactors
\citep{har07},  the spike recorded  in the  lunar melt  clasts implies
also the existence of a number of km-sized impactors. This enhancement
in  the km-sized  impactors may  explain why  the Earth  cratering for
$D>20$~km   seems   not   to   be  depleted   by   erosion   processes
(fig. \ref{mpf_neukum_fit}, left panel), as expected \citep{gri93}.

\subsection{On the early flux}

One of  the major -and mostly  debated- open issues is  related to the
cratering rate  in the  early times after  the formation of  the Moon.
Arguments have been  proposed in favor of a  rapid and smooth decrease
in the impact  rate from the formation of the  Moon till about 3.5~Gyr
ago  \citep{har81,   neu83},  followed  by  a   constant  impact  rate
\citep{neu83, neu94}.  On the other  hand, studies of impact melts led
some  researchers \cite[e.g.][]{ryd90,  sto01, coh00}  to  support the
idea of  an intense lunar  bombardment about 3.9~Gyr ago  which lasted
some 100-200~Myr.  From a dynamical point of view, such short phase of
intense bombardment  has been recently  explained in the  more general
context  of the  early stages  of the  evolution of  the  Solar System
\citep{gom05}.   This  scenario  has  also  found  some  observational
evidence in the work of \cite{str05}.\\
Despite these  works, the  LHB hypothesis has  not found  an unanimous
consensus  yet.  For  instance, \cite{har07}  raised doubts  about the
interpretation  of lunar  melts and  glasses data;  while \cite{neu94}
argued against  the LHB on the  basis of the smooth  behavior of their
lunar chronology curve.\\
If the intense LHB did take  place, it should have left some traces in
the  chronology  curve,  which  therefore  could  be  used  to  obtain
constraints  on the  early  stages of  the cratering.   Unfortunately,
there are only 5 measured regions having ages older than 3.85~Gyr, and
therefore it  is not easy to  draw firm conclusions.  One  of the most
important  regions  is  the   Nectaris  Basin.   \cite{neu94}  used  a
radiometric  age of  4.1~Gry.  The  resulting best  fit for  the lunar
chronology  curve is  therefore  very  close to  all  the old  regions
(within errors) and has a smooth behavior in early times, reflecting a
possible    smooth   decay    in   the    impact   rate    (see   fig.
\ref{mpf_neukum_fit}).   However, using the  new estimate  of 3.92~Gyr
for  Nectaris   Basin  proposed  by  \cite{sto01}   this  point  moves
considerably far  from the best  fit (this is  the age we used  in our
chronology).  This results in a  change of the slope of the chronology
curve, which in turn is in favor of the LHB.  We underline that in the
present analysis  we used the present NEO  size distribution. However,
it  is very  likely that  at the  time of  the LHB  the  impactor size
distribution had a different shape.  At the present stage of knowledge
a  firm  conclusion from  the  chronology  curve  cannot be  obtained.
Nevertheless, we  will add few considerations on  this important point
in the next section.

\section{A time-dependent MPF}\label{t-mpf}

In  this section we  develop a  time-dependent MPF.   In doing  so, we
 clearly  show the  versatility of  our  approach, as  opposed to  the
 conventional methods  (e.g.  the NPF), where  the production function
 is  generally   assumed  to  be  constant  over   time.   Modeling  a
 time-dependent MPF is quite complex, and in this section we will only
 pose the basis of the problem, deferring thorough investigations to a
 further analysis.   In general terms, there are  two possible reasons
 for such time-dependent  behavior: The first one is  related to a non
 stationary flux\footnote{We  stress once  again that in  our approach
 what  is relevant  is the  shape of  the impactor  size  and velocity
 distributions, since the absolute calibration of the model is done by
 the  reference  lunar regions.},  $\phi(d,v,t)$;  the  second one  to
 crater erasing processes.\\
A  motivation  for  studying the  non  stationary flux  is
provided  by the  dynamical processes  responsible for  the  flux.  In
early times (just  after the formation of the  terrestrial planets and
the Moon, some 4.5~Gyr ago) the inner Solar System was still populated
by  the  leftover   planetesimals,  which  were  rapidly  (10-100~Myr)
cleared-up by  the latest stages  of the accretion process.   A second
source  of impactors was  represented by  the highly  perturbed bodies
(e.g.   through the  {\it  sweeping resonance}  mechanism)  due to  the
formation and migration  of Jupiter. At the present  time, the flux is
mainly sustained by  the slow decay of MBAs  into the resonances.  All
these  three stages  have their  own characteristic  $\phi(d,v)$.  For
sake of  simplicity, here  we focus on  the last two  mechanisms.  The
second  process  is  size  dependent,  while the  first  one  is  not,
therefore  the shape of  the impactor  size distribution  changes over
time.   On  the other  hand,  the  impactor  velocity distribution  is
expected to remain approximately constant  over time, as in both cases
the transport  mechanism to the NEO  space is the  same.  Therefore we
may write $\phi(d,v,t)=h(d,t)f(d,v)$.\\
The  crater  erasing on  large  bodies  depends  mainly on  two  major
effects:  the   superposition  of   craters  and  the   local  jolting
\citep{obr06}. The first one basically accounts for the overlapping of
craters  as the  surface  crater density  increases.   The second  one
considers  the  erasing  of  craters  due  to  local  seismic  jolting
triggered  by the  formation of  large craters.   Both effects  can be
modeled by considering a variation in the number of craters in a given
size  bin \citep{obr06}.   Let ${\cal  E}$  be the  ratio between  the
actual number of craters considering  the erasing and the total number
neglecting the erasing, we can write:

\begin{equation}
{\rm MPF}(D,t)=\int_{D}^\infty \Phi(\tilde{D},t){\cal E}(\tilde{D},t)\,d\tilde{D}
\end{equation}

where  setting $t=0$  and  ${\cal E}=1$  reduces  to the  MPF used  in
previous sections  (see appendix for further details).   Here we limit
ourselves to  considering these  effects separately.  Let  us consider
first the  crater erasing mechanism,  assuming $\phi\equiv\phi(d,v,0)$
as   in  previous   sections.    The  results   are   shown  in   fig.
\ref{moon_iso}. It is clear that  as the age increases the MPF becomes
flatter  for  small dimensions.   This  is  the well-known  saturation
process of heavily cratered  surfaces.  This affects the derived $N_1$
values.   To study  possible variations  in the  rate of  impacts over
time,  we  adopted  the  following  procedure:  first  the  cumulative
distributions  for the calibration  regions were  fitted with  the MPF
corrected for the erasing, then the obtained $N_1$ was rescaled to the
total number of craters that  occurred in the region (i.e.  the actual
number of impactors).  In fig.  \ref{ce_mb} the effects of the erasing
on  the  chronology  are  shown.  Although  the  erasing  considerably
changes  the MPF  for the  old  ages (see  fig.  \ref{moon_iso}),  its
effects on the chronology are not important.\\
Let us now deal with changes  in the impactor size distribution and no
erasing  (${\cal E}\equiv1$).  Following  the previous  discussion, we
assumed that  before the end of  the LHB ($\sim3.8$~Gyr  ago) the size
distribution   $h(d)$    resembled   closely   that    of   the   main
belt\footnote{For  sake of simplicity,  we used  the present  MBA size
distribution although  it may have had  a different shape  in the past
\citep{bot05a, bot05b}.}  ($h_m$), while afterwards it becomes similar
to the present one, that is $h_n$.  This assumption is also encouraged
by the observation that $h_m$ provides a much better fit to the crater
distribution of the highlands and  Nectaris Basin than $h_n$ does (see
online  material).  With respect  to the  previous analysis  (see fig.
\ref{mpf_neukum_fit}),  the only difference  regards two  data points,
namely the highlands and Nectaris  Basin. For both regions we obtain a
lower $N_1$  value, with the  reduction being more pronounced  for the
highlands.   This  results  in  a  net  change in  the  slope  of  the
chronology in the  two oldest points, suggesting a  change in the rate
of impactors.   Therefore, if  the MBA size  distribution is  used for
earlier times on the basis of the improved fit with both the highlands
and Nectaris Basin, the chronology plot would suggest the existence of
the LHB.   These conclusions are weakly affected  by the uncertainties
in the model (size distribution, scaling law etc.)  because the MPF is
well defined for the large crater sizes relevant in this context. \\

\section{Mercury's chronology}\label{mercu}

The  impactor size and  velocity distributions  for Mercury  have been
computed in a similar way  as done for the Moon (figs \ref{merc_size},
\ref{merc_velocity}). We assumed density and strength profiles similar
to those  adopted for the  Moon.  The resulting  MPF is shown  in fig.
\ref{merc_scaling} (right panel), along  with the overplot of the NPF
for  Mercury  \citep{neuk01}.   The  two production  functions  nicely
overlap at the extremes, that is for $D<0.3$~km and $D>40$~km.  Inside
this  range, however,  the  two  curves are  quite  different, with  a
maximum  deviation  of nearly  a  factor  of  5 at  $D\sim3$~km  (fig.
\ref{merc_scaling}, left  panel).  The exact  behavior of the  MPF in
this range depends on the assumptions on the transition in the scaling
law   between   cohesive   soil    and   rock   (see   discussion   in
sect. \ref{scalinglaw}).  Nevertheless,  whatever reasonable choice is
made for the  scaling law parameters, our MPF  is remarkably different
from the NPF in this range.\\
As an  example, we  report here our  age determination of  the Chekhov
basin.   Crater counting was  performed on  the total  ejecta blanket,
over an area exceeding $4\cdot 10^4$~km$^2$. According to the analysis
of \cite{neuk01}, Chekhov  basin has a derived age  of 4.05~Gyr, which
makes it one of the oldest regions on Mercury.  In fig.  \ref{chekhov}
we report  our MPF-based age estimate,  obtained considering different
scenarios, namely  using the  NEO and MBA  size distributions.  In the
nominal case (i.e.  MPF=MPF$(D,0)$; ${\cal E}=1$), we obtain an age of
3.94~Gyr, which is slightly  younger than that found by \cite{neuk01}.
The  corresponding $\chi^2$ of  the best  fit is  0.8.  Note  that, in
comparison to  the NPF ($\chi^2$=1.5),  the MPF more closely  fits the
observed  crater cumulative distribution.   When considering  also the
crater erasing, the age becomes 4.05~Gyr ($\chi^2$=0.6), and the shape
of  the MPF  is basically  unchanged.   In the  case of  the MBA  size
distribution and no erasing, we obtain 3.97~Gyr ($\chi^2$=0.6). Notice
that in the case of MBA,  if we include crater erasing the MPF reaches
the saturation  prior to achieving the  best fit, therefore  it is not
possible to  derive a  precise age estimate.   In conclusion,  the MPF
shows a better  fit to the data points than the  NPF.  However, due to
the measurement errors and to  the limited crater size range observed,
it  is not  possible  to conclude  whether  the NEO  or  the MBA  size
distribution is more suitable for fitting the Chekhov basin.\\

\section{Conclusions}

In this paper a new model  for the chronology of planetary surfaces in
the inner Solar System has been introduced and applied to the Moon and
Mercury.  Concerning the Moon, the main findings are the following:

\begin{itemize}

\item The nominal MPF differs from the NPF for less than a factor of 2
  throughout  the observed range  of dimensions.  A similar  degree of
  agreement is obtained for the chronology;

\item  Despite the  good agreement  between MPF  and NPF,  there  is a
  systematic  misfit of the  assumed linear  branch of  the chronology
  function  with the  data  points  for ages  more  recent than  about
  0.4~Gyr.   It  is remarkable  that  the  Earth  craters (which  have
  $D>20$~km) are  aligned with the  Tycho, Cone and North  Ray craters
  (having $D<0.4$~km).   Although the misfit  may be partially  due to
  the uncertainties  present in the  model, other causes, such  as for
  instance a non-constant flux in recent times due to the formation of
  MB families (like Baptistina and Flora) might be possible;
  
\item  The   MPF  was  computed   for  two  different   impactor  size
 distributions, namely  the present NEO and MBA  populations.  In this
 way we were  able to study whether the  cratering production function
 changed over time.   Both the NEO and MBA-based MPFs  are able to fit
 the cumulative distributions having  ages $<3.9$~Gyr.  This is mainly
 due  to  the  fact  that  such  regions  have  craters  smaller  than
 $\sim10$~km, where the MPFs are very similar.  On the other hand, the
 oldest regions (highlands 4.35~Gyr; Nectaris Basin 3.92~Gyr) are best
 fitted  using the  MBA  size distribution.   This  suggests that  the
 impactor size distribution around 4~Gyr ago resembled the present MBA
 distribution,  while today  it  is  closer to  the  present NEO  size
 distribution. Alternatively,  it is also possible that  comets had in
 the  past  a non  negligible  contribution.   This  point deserves  a
 specific analysis and will be subject of a future work;
\end{itemize}

Concerning Mercury,  the MPF  differs from the  NPF in the  size range
from 0.3~km to 40~km.  The maximum difference -of about a factor of 5-
occurs  at  $\sim3$~km.   This  discrepancy is  therefore  potentially
important  for dating  purposes, in  particular when  measurements for
small  craters  will  become   available  (e.g.   from  MESSENGER  and
BepiColombo). Outside  this range, the  NPF and MPF are  very similar.
We show  the results of our model  for the case of  the Chekhov basin.
In the nominal case (MPF$(D,0)$,  ${\cal E}=1$) we obtained the age of
3.95~Gyr, which compares to the 4.05~Gyr of \cite{neuk01}.

\acknowledgments

We  are grateful  to  A.~Morbidelli for  helpful  discussions and  for
providing us with the results of the orbital integrations. We are also
grateful  to  R.~Wagner  for  providing  us with  the  original  lunar
cumulative distributions.  Thanks  to G.~Neukum for helpful discussion
and suggestions.   We also thank  W.~Bottke for providing us  with the
size distribution of NEOs and MBAs.  Finally, thanks to D.~De~Niem and
G.F.~Gronchi for helpful discussions.

\appendix

\section{The Model Production Function}

The cumulative distribution of  craters (the so-called {\it production
function})   can   be  evaluated   starting   from  the   differential
distribution  of  the incoming  flux,  namely  $\phi(d,v,t)$. Here  we
consider the general  problem of a non stationary  flux, therefore the
time  dependence  of the  function  $\phi$.   Each  impact produces  a
crater,  with   a  size  specified   by  the  scaling   law,  $D={\cal
S}(d,v,\vec{p})$.  Let $\Phi(D)$ be  the differential  distribution of
craters, namely the number of craters per unit of crater size per unit
surface. Such distribution is obtained by integrating the differential
distribution of impactors over the one-dimensional domain specified by
the  scaling  law,  namely  $\gamma_D: {\cal  S}(d,v,\vec{p})=D$.  Let
$\sigma$ be a parameter along the curve $\gamma_D$, we finally have:


\begin{equation}
\Phi(D,t)  = \int \phi(\gamma_D(\sigma),t)\biggl|\biggl|\frac{d\gamma_D}{d\sigma}(\sigma)\biggr|\biggr|d\sigma
\end{equation}



In the  presence of erasing  processes, the actual number  of measured
craters  in any given  size bin  $dD$ is  not equal  to the  number of
impactors. Let  ${\cal E}(D,t)$  be the ratio  of the final  number of
crater  erasing  included,  with   the  total  number  (i.e.   erasing
excluded), we finally obtain:

\begin{equation}
{\rm MPF}(D,t)=\int_{D}^\infty \Phi(\tilde{D},t){\cal E}(\tilde{D},t)\,d\tilde{D}
\end{equation}

Following the  work of  \cite{obr06}, the function  ${\cal E}$  can be
expressed as

\begin{equation}
{\cal E}(D,t)=\frac{1-e^{-A}}{A}
\end{equation}

where $A$ is the total  effective area of craters accumulated per unit
area.  Therefore  $A$ depends  on the age  since older regions  have a
large  number of  craters  accumulated  and hence  a  larger $A$  with
respect to younger regions.  $A$ also depends on crater diameter since
small craters  are more frequent than  larger ones. As  a results, the
counting of small craters on  old regions are strongly affected by the
erasing (see fig.  \ref{moon_iso}).  In calculating $A$, we considered
the  superposition of  craters and  the local  jolting,  and neglected
global jolting  and cumulative effects of seismic  shaking, since they
are not relevant  for large bodies like the  terrestrial planets.  The
parameters  involved in  the estimate  of $A$  have been  derived from
\cite{obr06}.  The implementation of the erasing processes in our code
has  been tested  accurately using  asteroid crater  counting (Gaspra,
Ida, etc). Our results  agree with those published.  Nevertheless, the
purpose  of the  present analysis  is simply  to show  the first-order
effects of  the erasing, and a  detailed evaluation of  the erasing on
planetary-sized bodies is left for further analysis.

\clearpage

\begin{deluxetable}{r|ccc}
\tabletypesize{\scriptsize}
\tablecaption{Description of the lunar and terrestrial calibration
  regions  used in this work.}
\tablewidth{0pt}
\tablehead{
\colhead{Region} & \colhead{$N_1$ (NEO)$^\dag$} & \colhead{$N_1$ (MBA)$^\dag$} & \colhead{Age$^\ddag$} \\
                 &  (km$^{-2}$)     &  (km$^{-2}$)                 &  (Gyr)
}
\startdata
Highlands$^a$                             & $7.851 \cdot 10^{-1}$          &$2.018 \cdot 10^{-1}$          & 4.35 \\
Nectaris Basin                            & $1.327 \cdot 10^{-1}$          &$6.648 \cdot 10^{-2}$          & 3.92 \\
Descartes Formation$^{b,s}$               & $2.490 \cdot 10^{-2}$          &$2.509 \cdot 10^{-2}$          & 3.92 \\
Imbrium Apennines$^c$                     & $1.968 \cdot 10^{-2}$          &$1.931 \cdot 10^{-2}$          & 3.85\\
Fra Mauro Formation                       & $2.595 \cdot 10^{-2}$          &$2.672 \cdot 10^{-2}$          & 3.85\\
Mare Tranquillitatis (old)$^{d,s}$        & $1.836 \cdot 10^{-2}$          &$1.832 \cdot 10^{-2}$          & 3.80\\
Taurus Littrow Mare$^s$                   & $1.579 \cdot 10^{-2}$          &$1.585 \cdot 10^{-2}$          & 3.70 \\
Mare Tranquillitatis (young)$^{d,s}$      & $9.300 \cdot 10^{-3}$          &$9.357 \cdot 10^{-3}$          & 3.58 \\
Mare Fecunditatis$^s$                     & $3.234 \cdot 10^{-3}$          &$3.257 \cdot 10^{-3}$          & 3.41 \\
Mare Imbrium$^s$                          & $5.468 \cdot 10^{-3}$          &$5.526 \cdot 10^{-3}$          & 3.30 \\
Mare Crisium                              & $2.335 \cdot 10^{-3}$          &$2.377 \cdot 10^{-3}$          & 3.22 \\
Oceanus Procellarum                       & $3.683 \cdot 10^{-3}$          &$3.695 \cdot 10^{-3}$          & 3.15\\
Copernicus Crater (cont. ejecta)$^e$      & $1.321 \cdot 10^{-3}$          &$1.337 \cdot 10^{-3}$          & 0.80\\
Copernicus Crater (crater floor)$^e$      & $1.348 \cdot 10^{-3}$          &$1.343 \cdot 10^{-3}$          & 0.80\\
Terrestrial Phanerozoic craters$^f$       & $1.267 \cdot 10^{-3}$          &$7.655 \cdot 10^{-4}$          & 0.375\\
Terrestrial Phanerozoic craters (young)$^g$ & $3.835 \cdot 10^{-4}$        &$2.195 \cdot 10^{-4}$          & 0.120\\
Tycho crater (cont. ejecta)$^h$           & $3.391 \cdot 10^{-4}$          &$3.401 \cdot 10^{-4}$          & 0.109\\
Tycho crater (cont. ejecta)$^i$           & $1.644 \cdot 10^{-4}$          &$1.712 \cdot 10^{-4}$          & 0.109\\
North Ray crater$^l$                      & $1.389 \cdot 10^{-4}$          &$1.421 \cdot 10^{-4}$          & 0.053\\
Cone Crater$^l$                           & $6.970 \cdot 10^{-5}$          &$7.131 \cdot 10^{-5}$          & 0.025  \\
\enddata
\tablecomments{Where  not   explicitly  quoted,  the   data  are  from
\cite{neu83} and references therein.  The plots showing the fits of the
calibration regions are reported in the online material.}
\tablenotetext{\dag}{The derived  $N_1$ values both using  the NEO and
  MBA size distributions.}
\tablenotetext{\ddag}{Radiometric ages for  the lunar regions are from
 table VI of \cite{sto01}.  The ages of  the terrestrial craters
 are from \cite{gri79}. }
\tablenotetext{a}{All  craters   and  basins.}   
\tablenotetext{b}{The counting  for Descartes Formation  is limited to
dimensions  smaller  than  1.2~km,  while that  used  in  \cite{neu83}
extends to larger sizes.}
\tablenotetext{c}{The crater  counting for Imbrium  Apennines (used in
  this  work) is  a sub  sample of  the crater  counting  from Imbrium
  Basin, used by \cite{neu83}.}
\tablenotetext{d}{Original   data  from   \cite{gre70}.}
\tablenotetext{e}{For  the  Copernicus  crater  we have  two  distinct
measurements  corresponding to  the  crater floor  and the  continuous
ejecta blanket.  In fig.~31 of \cite{neu83}, these  sets were probably
merged into a single distribution.   We decided to keep these two sets
separate,    and    to    provide    individual   fits    (see    fig.
\ref{mpf_neukum_fit} and the online material).  }
\tablenotetext{f}{Terrestrial  Phanerozoic  craters  are derived  from
\cite{gri79},  and  updated according  to  the  Earth Impact  Database
(http://www.unb.ca/passc/ImpactDatabase/).     Only    large   craters
($D>20$~km)  on the  North American  and Euroasian  cratons  have been
considered. For  sake of  completeness we report  the list  of craters
used. In the North  American craton: Beaverhead, Carswell, Charlevoix,
Clearwater  East,  Clearwater  West,  Haughton,  Manicouagan,  Manson,
Mistastin,  Presquile,  Saint  Martin,  Slate  Islands,  Steen  River,
Sudbury.  In  the  Eurasian  craton:  Boltysh,  Kamensk,  Keurusselka,
Lappajarvi, Puchezh-Katunki, Siljan.}
\tablenotetext{g}{Only  large ($D>20$~km) and  young ($<120$~Myr)
  terrestrial   craters.  They   are:   Carswell,  Haughton,   Manson,
  Mistastin,    Steen    River,    Boltysh,   Kamensk,    Lappajarvi.}
 \tablenotetext{h,i}{Two distinct measurements  were available for the
  Tycho  crater.  Both  correspond to  the continuous  ejecta:  in (h)
  small craters were counted, while in (i) large craters were counted.
  In fig.~31  of \cite{neu83}  these files were  merged into  a single
  distribution.  We  decided to keep  these two sets separate,  and to
  provide  individual fits  (see fig.   \ref{mpf_neukum_fit}).  Notice
  that  \cite{neu83}, also  reports  4 measured  points  in the  range
  0.5-1.1~km that were not available to us. }
  \tablenotetext{l}{Original   data  from \cite{moo80}.} 
\tablenotetext{s}{These  are  relatively  old  regions  having  crater
  counting  extending to  small sizes  ($<1$~km) and  thus potentially
  affected  by secondary  craters.  In order  to  check whether  these
  regions affect  the final MPF-based chronology, we  also performed a
  best  fit  excluding  these  points.   The results  of  the  fit  is
  basically unchanged (see text for further details.)}
%
\label{table1}
\end{deluxetable}

\clearpage

\begin{figure}
\includegraphics[angle=-90,scale=.50]{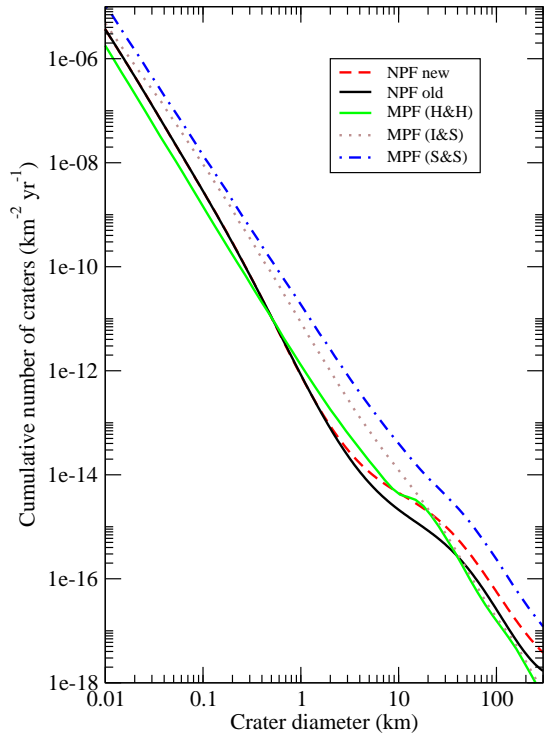}
\caption{Comparison of  the MPF obtained with  different scaling laws,
 indicated  by  H\&H, I\&S,  S\&S  (see  section \ref{scalinglaw}  for
 details).  The old and new  NPFs are also shown. }
\label{moon_scaling}
\end{figure}

\begin{figure}
\includegraphics[angle=-90,scale=.50]{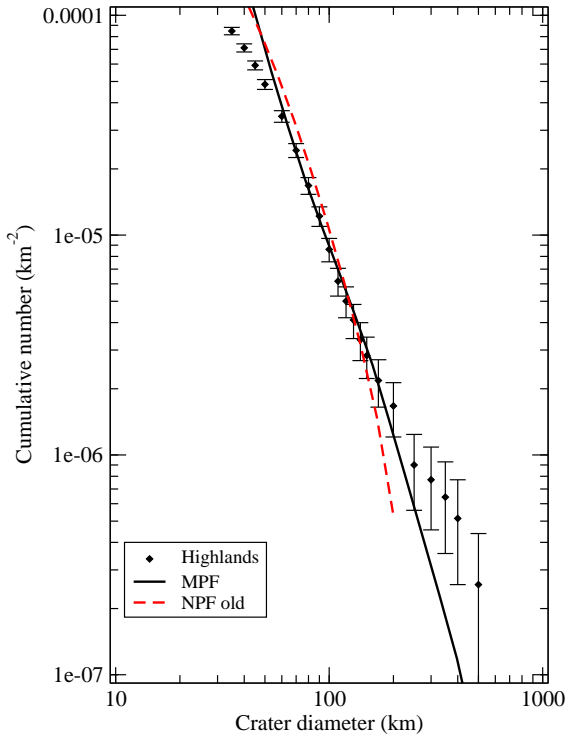}
\caption{The plot shows the best fit of the crater-size cumulative
  distribution for the lunar highlands with both  the MPF and the NPF. }
\label{highlands}
\end{figure}

\begin{figure}
\includegraphics[angle=-90,scale=.35]{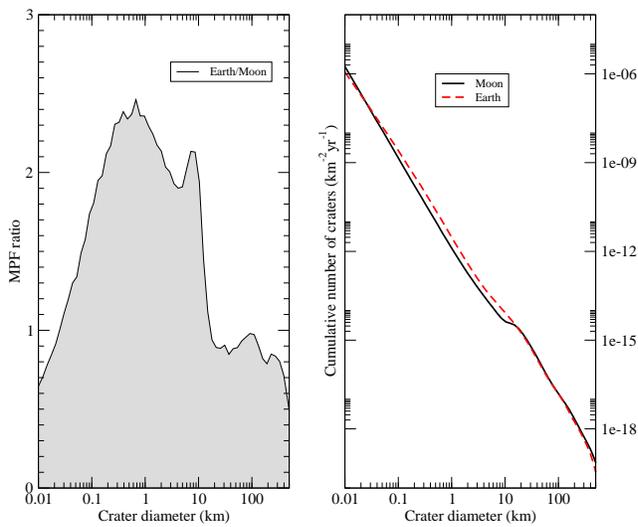}
\caption{Detailed comparison of the lunar and terrestrial MPFs (using
  the H\&H scaling law).}
\label{moon_earth}
\end{figure}

%
%

\begin{figure}
\includegraphics[angle=-90,scale=.55]{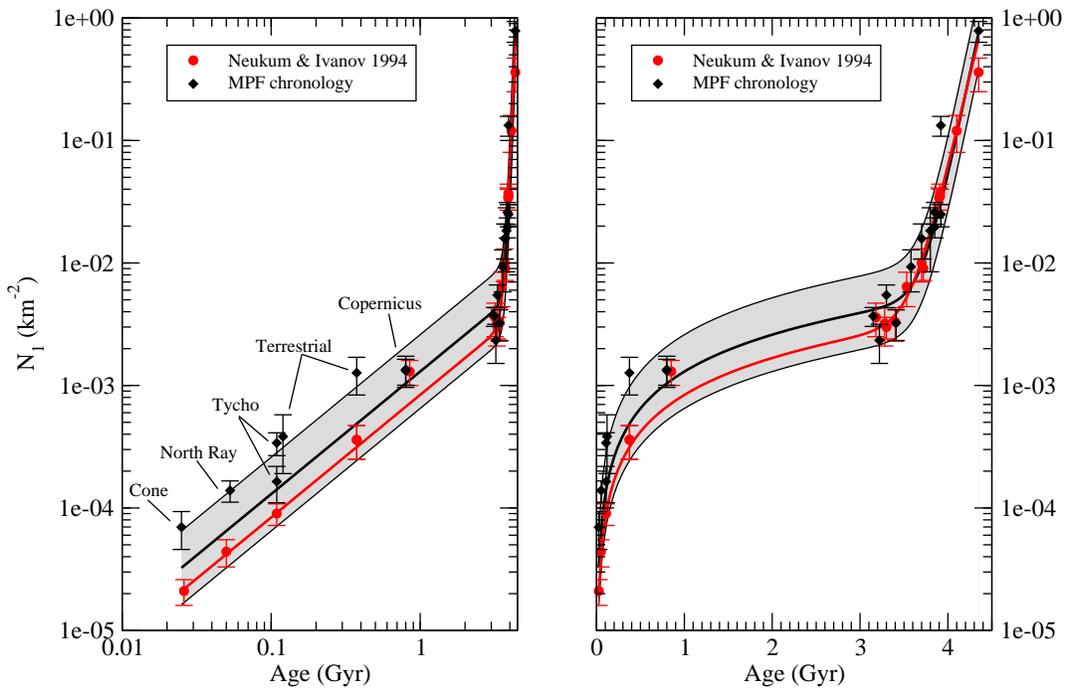}
\caption{The plot  shows the $N_1$ values for  the calibration regions
  obtained  using the  MPF and  those obtained  by  \cite{neu94}.  The
  solid curves,  namely the chronology curves, represent  the best fit
  of the two sets of  data points.  The shadowed regions encompass a
  factor of $\pm 2$ around the MPF chronology curve.}
\label{mpf_neukum_fit}
\end{figure}


\begin{figure}
\includegraphics[angle=-90,scale=.50]{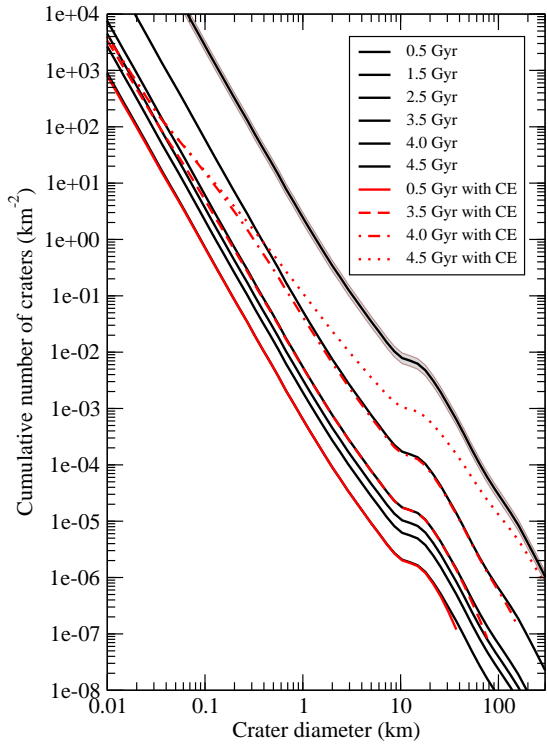}
\caption{Moon isochrones  obtained with MPF$(D,0)$,  and including the
 crater-erasing (CE) mechanism.  The effect of longitude variations is
 also shown by  the shadowed area  for the  4.5~Gyr curve (with no
 erasing).  The  expected longitudinal asymmetry in  the impactor flux
 has  been modeled  using eq.~1  of \cite{mor05}.  The impact  of such
 analysis on the $N_1$  values is negligible, therefore the chronology
 is not affected.  }
\label{moon_iso}
\end{figure}

\begin{figure}
\includegraphics[angle=-90,scale=.55]{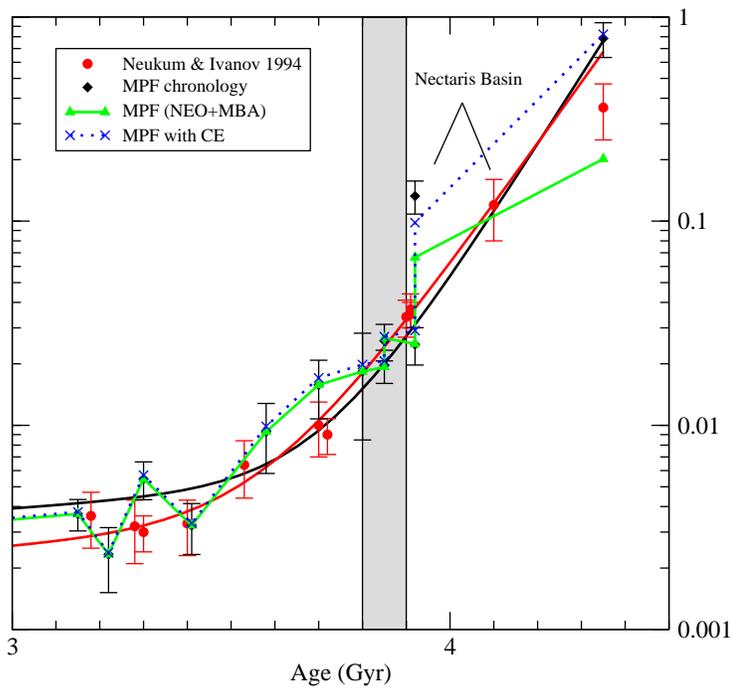}
\caption{In this figure the effects  of using the size distribution of
  MBAs and  the crater-erasing  (CE) mechanism are  shown for  the old
  regions alone, since the young  ones are not affected.  The vertical
  strip  corresponds approximately  the  LHB event.   Notice that  for
  Nectaris  basin  the age  from  \cite{sto01}  adopted  in this  work
  differs  from the  one in  \cite{neu94}.  Moreover,  the age  of the
  highlands is known with large  uncertainty and may vary in the range
  4.2-4.4~Gyr. Here we adopted the values used by \cite{neu94}.}
\label{ce_mb}
\end{figure}

\begin{figure}
\includegraphics[angle=-90,scale=.50]{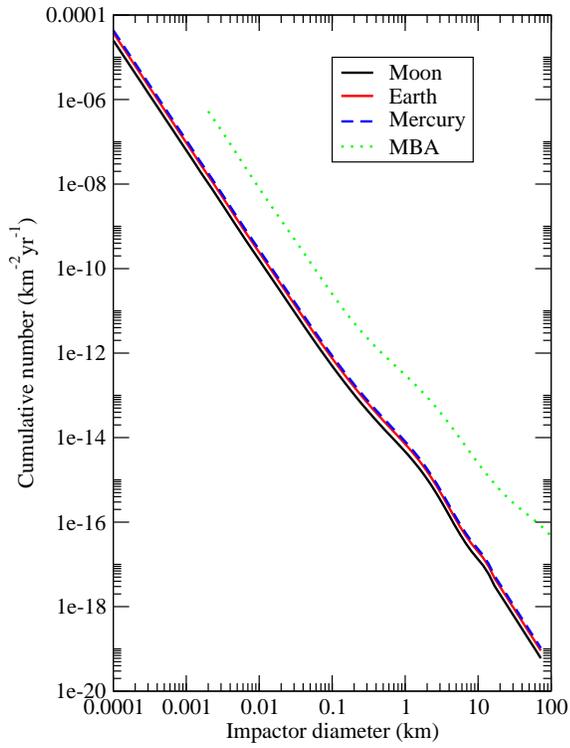}
\caption{The plot shows the model cumulative distribution of impactors
 for Mercury,  in comparison with  that of the Earth-Moon  system. The
 MBA size distribution is also shown.}
\label{merc_size}
\end{figure}

\begin{figure}
\includegraphics[angle=-90,scale=.35]{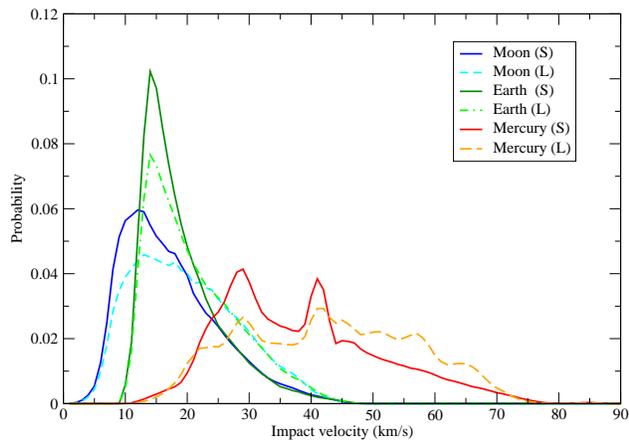}
\caption{Impactor  velocity distribution  for Mercury,  in comparison
with the Earth-Moon system (see also fig. \ref{moon_velocity}).}
\label{merc_velocity}
\end{figure}

\begin{figure}
\includegraphics[angle=-90,scale=.35]{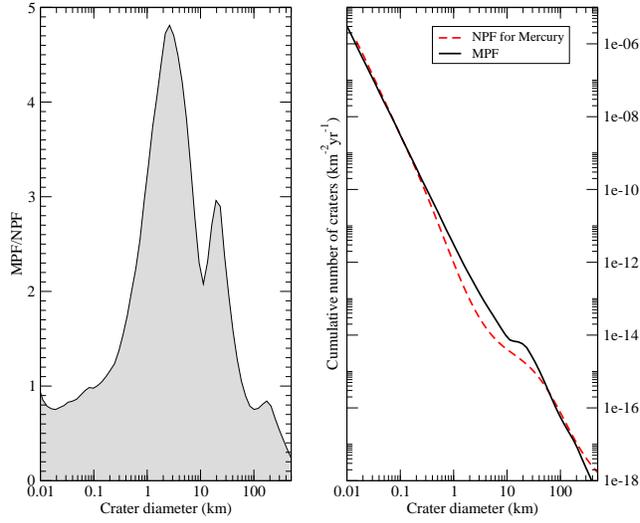}
\caption{MPF  for  Mercury,  in  comparison  with  the  NPF.  The  two
  production  functions are in  disagreement in  the range  from about
  0.1~km to 40~km.}
\label{merc_scaling}
\end{figure}

\begin{figure}
\includegraphics[angle=-90,scale=.50]{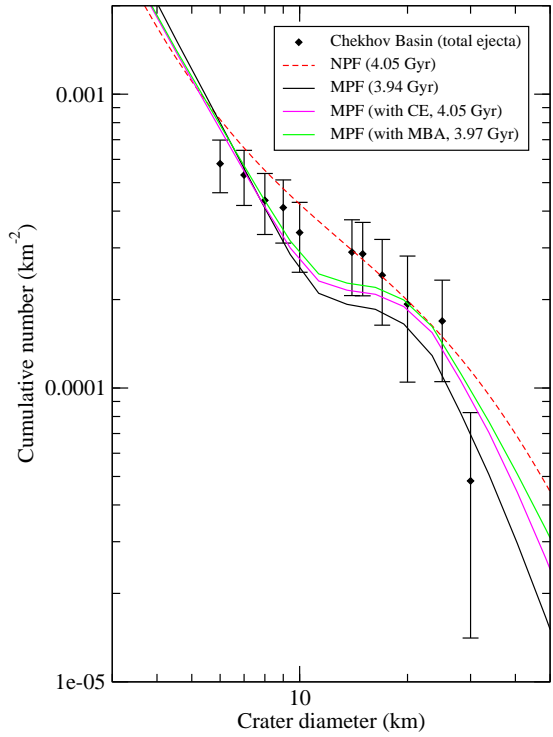}
\caption{MPF  best fit  of Chekhov  basin  and derived  age.  We  also
  present the best  fit using the MBA size  distribution and crater
  erasing (see section \ref{t-mpf}).}
\label{chekhov}
\end{figure}


\begin{thebibliography}{}


\bibitem[Anderson (2007)]{and07} Anderson,  D.~L.\ 2007, New Theory of
the Earth. Cambridge University press, 380 pp.

\bibitem[Asphaug et al.(1996)]{asp96} Asphaug, E., Moore, 
J.~M., Morrison, D., Benz, W., Nolan, M.~C., 
\& Sullivan, R.~J.\ 1996, Icarus, 120, 158 

\bibitem[Birck   \&   All{\`e}gre(1978)]{bir78}   Birck,
J.~L., \& All{\`e}gre, C.~J.\ 1978, Physics of the Earth and Planetary
Interiors, 16, 10

\bibitem[Bogard(1995)]{bog95} Bogard, D.\ 1995, Meteoritics, 
30, 244 

\bibitem[Bottke et al.(2007)]{bot07} Bottke, W.~F., 
Vokrouhlick{\'y}, D., \& Nesvorn{\'y}, D.\ 2007, \nat, 449, 48 

\bibitem[Bottke et al.(2005a)]{bot05a} Bottke, W.~F., Durda, 
D.~D., Nesvorn{\'y}, D., Jedicke, R., Morbidelli, A., Vokrouhlick{\'y}, D., 
\& Levison, H.\ 2005a, Icarus, 175, 111 

\bibitem[Bottke et al.(2005b)]{bot05b} Bottke, W.~F., Durda, 
D.~D., Nesvorn{\'y}, D., Jedicke, R., Morbidelli, A., Vokrouhlick{\'y}, D., 
\& Levison, H.~F.\ 2005b, Icarus, 179, 63 

\bibitem[Bottke et al.(2002)]{bot02} Bottke, W.~F., 
Morbidelli, A., Jedicke, R., Petit, J.-M., Levison, H.~F., Michel, P., 
\& Metcalfe, T.~S.\ 2002, Icarus, 156, 399 

\bibitem[Bottke et al.(2000)]{bot00} Bottke, W.~F., Jedicke, 
R., Morbidelli, A., Petit, J.-M., \& Gladman, B.\ 2000, Science, 288,
2190 

\bibitem[Brown et al.(2002)]{bro02} Brown, P., Spalding, 
R.~E., ReVelle, D.~O., Tagliaferri, E., 
\& Worden, S.~P.\ 2002, \nat, 420, 294 

\bibitem[Boyce et al.(1977)]{boy77} Boyce, J.~M., Schaber, 
G.~G., \& Dial, A.~L., Jr.\ 1977, \nat, 265, 38 

\bibitem[Cohen et al.(2000)]{coh00} Cohen, B.~A., Swindle, 
T.~D., \& Kring, D.~A.\ 2000, Science, 290, 1754 

\bibitem[Cohen   et   al.(2005)]{coh05}  Cohen,   B.~A.,
Swindle,  T.~D.,  \& Kring,  D.~A.\  2005,  Meteoritics and  Planetary
Science, 40, 755

\bibitem[Culler et al.(2000)]{cul00} Culler, T.~S., Becker, 
T.~A., Muller, R.~A., \& Renne, P.~R.\ 2000, Science, 287, 1785 


\bibitem[Farinella et al.(1998)]{far98} Farinella, P., 
Vokrouhlicky, D., \& Hartmann, W.~K.\ 1998, Icarus, 132, 378 

\bibitem[Fernandes  \&  Burgess(2005)]{fer05} Fernandes,
V.~A., \& Burgess, R.\ 2005, \gca, 69, 4919

\bibitem[French   (1998)]{fre98}French,   B.~M.\  1998.   Traces   of
Catastrophe:  A Handbook of  Shock-Metamorphic Effects  in Terrestrial
Meteorite  Impact  Structures. LPI  Contribution  No.  954, Lunar  and
Planetary Institute, Houston. 120 pp.

\bibitem[Glass(1990)]{gla90} Glass, B.~P.\ 1990, 
Tectonophysics, 171, 393 

\bibitem[Gomes et al.(2005)]{gom05} Gomes, R., Levison, 
H.~F., Tsiganis, K., \& Morbidelli, A.\ 2005, \nat, 435, 466 

\bibitem[Greeley 
\& Gault(1970)]{gre70} Greeley, R., \& Gault, D.~E.\ 1970, Moon, 2, 10 


\bibitem[Grieve \& Dence(1979)]{gri79} Grieve, R.~A.~F.,
\& Dence, M.~R.\ 1979, Icarus, 38, 230

\bibitem[Grieve(1993)]{gri93}  Grieve,  R.~A.~F.\  1993,
Vistas in Astronomy, 36, 203


\bibitem[Halliday  et  al.(1996)]{hal96}  Halliday,  I.,
Griffin, A.~A.,  \& Blackwell, A.~T.\ 1996,  Meteoritics and Planetary
Science, 31, 185

\bibitem[Hartmann  et   al.(1981)]{har81}Hartmann,  W.K.,  Strom,  R.,
  Weidenschilling, S.,  Blasius, K.,  Woronow, A., Dence,  M., Grieve,
  R.,  Diaz,   J.,  Chapman,  C.,  Shoemaker,  E.,   Jones  K.,  1981.
  Chronology  of   planetary  volcanism  by   comparative  studies  of
  planetary  cratering.  In  Basaltic  Volcanism  on  the  Terrestrial
  Planets, pp. 1050-1129.  Basaltic Volcanism Study Project.

\bibitem[Hartmann et al.(2007)]{har07} Hartmann, W.~K., 
Quantin, C., \& Mangold, N.\ 2007, Icarus, 186, 11 

\bibitem[Holsapple 
\& Housen(2007)]{hol07} Holsapple, K.~A., \& Housen, K.~R.\ 2007, Icarus, 191, 586 

\bibitem[Koeberl et al.(1996)]{koe96} Koeberl, C., Poag, 
C.~W., Reimold, W.~U., \& Brandt, D.\ 1996, Science, 271, 1263 


\bibitem[Ivanov(2001)]{iva01} Ivanov, B.~A.\ 2001, Space 
Science Reviews, 96, 87 

\bibitem[Ivanov(2006)]{iva06} Ivanov, B.~A.\ 2006, Icarus, 
183, 504 

\bibitem[Ivanov et al.(2001)]{iva01} Ivanov, B.~A., Neukum, 
G., \& Wagner, R.\ 2001, Astrophysics and Space Science Library, 261, 1 

\bibitem[Ivanov et al.(2002)]{iva02} Ivanov, B.~A., Neukum, 
G., Bottke, W.~F., Jr., \& Hartmann, W.~K.\ 2002, Asteroids III, 89 

\bibitem[Marchi  et al.(2005)]{mar05} Marchi,  S., Morbidelli,  A., \&
Cremonese, G.\ 2005, \aap, 431, 1123

\bibitem[Moore et al.(1980)]{moo80} Moore, H.~J., Boyce,
J.~M., \& Hahn, D.~A.\ 1980, Moon and Planets, 23, 231

\bibitem[Morbidelli  et   al.(2002)]{mor02}  Morbidelli,  A.,  Bottke,
W.~F., Jr., Froeschl{\'e}, C., \& Michel, P.\ 2002, Asteroids III, 409

\bibitem[Morbidelli   \&  Gladman(1998)]{mor98}  Morbidelli,   A.,  \&
Gladman, B.\ 1998, Meteoritics and Planetary Science, 33, 999

\bibitem[Morota et al.(2005)]{mor05} Morota, T., Ukai, T., 
\& Furumoto, M.\ 2005, Icarus, 173, 322

\bibitem[Nesvorn{\'y} et al.(2007)]{nes07} Nesvorn{\'y},
D.,   Vokrouhlick{\'y},   D.,   Bottke,   W.~F.,   Gladman,   B., 
H{\"a}ggstr{\"o}m, T.\ 2007, Icarus, 188, 400

\bibitem[Neukum  \& Ivanov(1994)]{neu94} Neukum,  G., \&
Ivanov, B.~A.\ 1994, Hazards Due to Comets and Asteroids, 359

\bibitem[Neukum(1983)]{neu83}     Neukum,      G.,     PhD     Thesis,
Meteoritenbombardement und  Datierung Planetarer Oberflaechen, Munich,
Feb. 1983 p 1-186

\bibitem[Neukum et al.(2001)]{neu01} Neukum, G., Ivanov, 
B.~A., \& Hartmann, W.~K.\ 2001, Space Science Reviews, 96, 55 

\bibitem[Neukum \& Horn(1976)]{neu76} Neukum, G., \& Horn, P.\ 1976,
  Moon, 15, 205 

\bibitem[Neukum   et  al.(2001b)]{neuk01}   Neukum,  G.,   Oberst,  J.,
Hoffmann, H., Wagner, R., \& Ivanov, B.~A.\ 2001b, \planss, 49, 1507


\bibitem[O'Brien et al.(2006)]{obr06} O'Brien, D.~P., 
Greenberg, R., \& Richardson, J.~E.\ 2006, Icarus, 183, 79 

\bibitem[Pike(1980)]{pik80} Pike, R.~J.\ 1980, Icarus, 43, 1 


\bibitem[Schmidt \& Housen(1987)]{sch87} Schmidt, R.~M.,
\& Housen,  K.~R.\ 1987, International Journal  of Impact Engineering,
5, 543

\bibitem[Shoemaker  et al.(1990)]{sho90}Shoemaker, E.M.,  Wolfe, R.F.,
Shoemaker, C.S., 1990. Asteroid and  comet flux in the neighborhood of
Earth. Geological Society of America Special Paper 247


\bibitem[St{\"o}ffler       \&       Ryder(2001)]{sto01}
St{\"o}ffler, D., \& Ryder, G.\ 2001, Space Science Reviews, 96, 9

\bibitem[Strom et al.(2005)]{str05} Strom, R.~G., Malhotra, 
R., Ito, T., Yoshida, F., \& Kring, D.~A.\ 2005, Science, 309, 1847 


\bibitem[Stuart \& Binzel(2004)]{stu04} Stuart, J.~S., \& Binzel, R.~P.\ 2004, Icarus, 170, 295 

\bibitem[Ryder(1990)]{ryd90} Ryder, G.\ 1990, EOS 
Transactions, 71, 313 


\bibitem[Werner et al.(2002)]{wer02} Werner, S.~C., Harris, 
A.~W., Neukum, G., \& Ivanov, B.~A.\ 2002, Icarus, 156, 287 

\bibitem[Wilhelms(1976)]{wil76} Wilhelms, D.~E.\ 1976, Lunar 
and Planetary Science Conference, 7, 2883 


\end{thebibliography}
\end{document}